\pdfoutput=1
\documentclass[useAMS,usenatbib]{mn2e}
\usepackage[pdftex]{graphicx}
\usepackage{amssymb,amsmath}
\usepackage{wasysym}
\usepackage{url}
\usepackage{lineno}

\newcommand{\HMS}[3]{$#1^{\mathrm{h}}#2^{\mathrm{m}}#3^{\mathrm{s}}$}
\newcommand{\DMS}[3]{$#1^\circ #2' #3''$}
\newcommand{\degree}{$^\circ$}
\newcommand{\etacar}{Eta Carinae}
\newcommand{\zetastd}{$\zeta_{\mathrm{std}}$}
\newcommand{\UNITS}[1]{\,\mathrm{#1}}
\newcommand{\hess}{H.E.S.S.}
\newcommand{\g}{$\gamma$}
\newcommand{\order}{$\mathcal{O}$}
\newcommand{\TwoFGL}{2FGL~J1045.0--5941}

\title{\hess\ observations of the Carina nebula and its enigmatic colliding wind 
binary \etacar}

\author[HESS Collaboration]{\normalsize HESS Collaboration,
 A.~Abramowski,$^{1}$
 F.~Acero,$^{2}$
 F.~Aharonian,$^{3,4,5}$
 A.G.~Akhperjanian,$^{6,5}$
 G.~Anton,$^{7}$
 A.~Balzer,$^{7}$
 \newauthor \normalsize
 A.~Barnacka,$^{8,9}$
 Y.~Becherini,$^{10,11}$
 J.~Becker,$^{12}$
 K.~Bernl\"ohr,$^{3,13}$
% ...
\\
$^1$
Universit\"at Hamburg, Institut f\"ur Experimentalphysik, Luruper Chaussee 149, D 22761 Hamburg, Germany\\
$^2$
Laboratoire Univers et Particules de Montpellier, Universit\'e Montpellier 2, CNRS/IN2P3,  CC 72, Place Eug\`ene Bataillon,\\F-34095 Montpellier Cedex 5, France\\
$^3$
Max-Planck-Institut f\"ur Kernphysik, P.O. Box 103980, D 69029 Heidelberg, Germany\\
$^4$
Dublin Institute for Advanced Studies, 31 Fitzwilliam Place, Dublin 2, Ireland\\
$^5$
National Academy of Sciences of the Republic of Armenia, Yerevan\\
$^6$
Yerevan Physics Institute, 2 Alikhanian Brothers St., 375036 Yerevan, Armenia\\
$^7$
Universit\"at Erlangen-N\"urnberg, Physikalisches Institut, Erwin-Rommel-Str. 1, D 91058 Erlangen, Germany\\
$^8$
Nicolaus Copernicus Astronomical Center, ul. Bartycka 18, 00-716 Warsaw, Poland\\
$^9$
CEA Saclay, DSM/IRFU, F-91191 Gif-Sur-Yvette Cedex, France\\
$^{10}$
APC, AstroParticule et Cosmologie, Universit\'{e} Paris Diderot, CNRS/ IN2P3,CEA/ lrfu, Observatoire de Paris,\\Sorbonne Paris Cit\'{e}, 10, rue Alice Domon et L\'{e}onie Duquet, 75205 Paris Cedex 13, France\\
$^{11}$
Laboratoire Leprince-Ringuet, Ecole Polytechnique, CNRS/IN2P3, F-91128 Palaiseau, France\\
$^{12}$
Institut f\"ur Theoretische Physik, Lehrstuhl IV: Weltraum und Astrophysik, Ruhr-Universit\"at Bochum, D 44780 Bochum, Germany\\
$^{13}$
Institut f\"ur Physik, Humboldt-Universit\"at zu Berlin, Newtonstr. 15, D 12489 Berlin, Germany\\
% ...
}

\begin{document}

\date{Accepted 2012 April 25. Received 2012 April 24; in original form
  2012 March 27}
\pagerange{\pageref{firstpage}--\pageref{lastpage}} \pubyear{2012}

\maketitle

\label{firstpage}

\begin{abstract}
  The massive binary system \etacar\ and the surrounding HII complex,
  the Carina Nebula, are potential particle acceleration sites from
  which very-high-energy (VHE; $E\ge100$\,GeV) \g-ray emission could
  be expected. This paper presents data collected during VHE \g-ray
  observations with the \hess\ telescope array from 2004 to 2010,
  which cover a full orbit of \etacar. In the 33.1-hour data set no
  hint of significant \g-ray emission from \etacar\ has been found and
  an upper limit on the \g-ray flux of
  $7.7\times10^{-13}\UNITS{ph\,cm^{-2}\,s^{-1}}$ (99\% confidence
  level) is derived above the energy threshold of 470\,GeV. Together
  with the detection of high-energy (HE; 0.1\,GeV $\le E \le$
  100\,GeV) \g-ray emission by the \textit{Fermi}-LAT up to 100\,GeV,
  and assuming a continuation of the average HE spectral index into
  the VHE domain, these results imply a cut-off in the \g-ray spectrum
  between the HE and VHE \g-ray range. This could be caused either by
  a cut-off in the accelerated particle distribution or by severe
  \g-\g\ absorption losses in the wind collision region. Furthermore,
  the search for extended \g-ray emission from the Carina Nebula
  resulted in an upper limit on the \g-ray flux of
  $4.2\times10^{-12}\UNITS{ph\,cm^{-2}\,s^{-1}}$ (99\% confidence
  level). The derived upper limit of $\sim23$ on the cosmic-ray
  enhancement factor is compared with results found for the old-age
  mixed-morphology supernova remnant W~28.
\end{abstract}


\maketitle

\section{Introduction}

The Carina Nebula is one of the largest and most active HII regions in
our Galaxy and a place of ongoing star formation. It is located at a
distance of $\sim2.3$\,kpc and harbours eight open stellar clusters
with more than 66 O-type stars, 3 Wolf-Rayet stars and the Luminous
Blue Variable (LBV) \etacar\
\citep{Car:Feinstein95,Car:Smith06,Car:SmithBrooks08}. The existence
of a $\sim10^6$-year-old neutron star indicates past supernova (SN)
activity in the Carina complex
\citep{Car:Hamaguchi09,Car:Pires09}. The age estimates of the member
clusters Trumpler 14, 15 and 16 vary significantly, with an age
spread of $\sim2$\,Myr to $\sim8$\,Myr, indicating several past
episodes of star-formation in the northern region; more recent star
formation is going on in the southern part of the nebula
\citep[see][and references therein]{Car:Preibisch2011a}. Extended
X-ray emission has been reported by, e.g., \citet{Car:Hamaguchi07}
based on observations with Suzaku, supplemented by XMM-Newton
\citep{Car:Ezoe08} and Chandra \citep{Car:Townsley11}
observations. These authors found a very low nitrogen-to-oxygen ratio,
that, in addition to the presence of a neutron star, suggests that the
diffuse plasma originates in one or several unrecognised supernova
remnants, in particular in the area surrounding \etacar. The emission
may also be attributed to stellar winds from massive stars. In their
$\sim1.4$ sq.deg. survey of the diffuse X-ray emission,
\citet{Car:Townsley11} also found evidence for a significant
contribution due to charge exchange. This mechanism would originate in
a contact layer between the hot plasma and the cold molecular clouds.

\etacar, a member star of Trumpler\,16 (Tr\,16), is one of the most
peculiar objects in our Galaxy, whose environment shows traces of
massive eruptions that occurred in past epochs. A giant outburst in
the 1840s (known as the Great Eruption) and a smaller outburst in the
1890s produced the Homunculus and little Homunculus Nebulae,
respectively \citep[see e.g.][]{EtaCar:Ishibashi03}. The material
expelled from the central star in the Great Eruption has a combined
mass of $\sim12\,M_{\odot}$ and moves outwards at an average speed of
$\sim650$\,km\,s$^{-1}$, implying a kinetic energy of roughly
$(4-10)\times10^{49}$\,erg
\citep{EtaCar:Smith03}. \citet{EtaCar:Smith08} found material that is
moving ahead of the expanding Homunculus Nebula at speeds of
$3500-6000$\,km\,s$^{-1}$, which doubles the estimate of the kinetic
energy of the giant outburst. For a long time it was believed that the
central object, \etacar, is a single, hypergiant LBV star -- one of
only very few found in the Galaxy \citep[see e.g.][]{LBV:Clark05}.
However, observations now suggest \etacar\ to be composed of a massive
LBV star and an O- or B-type companion star
\citep{EtaCar:Hillier01,EtaCar:Pittard02}. The present-day period of
the binary has been estimated to $P_{\rm orb} = 2022.7\pm1.2$ days
\citep{EtaCar:Damineli08}, its eccentricity to be $e \sim0.9$
\citep{EtaCar:Nielsen07} and the semi-major axis to be
$a=16.64\UNITS{AU}$ \citep{EtaCar:Hillier01}. The LBV star has a very
high mass loss rate of $\dot{M}_1 \geq
5\times10^{-4}\,M_{\odot}\,\UNITS{yr}^{-1}$,
\citep{EtaCar:Hillier01,EtaCar:Parkin09} and a terminal wind velocity
of $v_1 \sim(500 - 700)\UNITS{km\,s^{-1}}$, the companion star has a
thin, fast wind ($\dot{M_2}
\sim1.0\times10^{-5}~M_{\odot}\UNITS{yr}^{-1}$ and $v_2
\sim3000\UNITS{km\,s^{-1}}$, \citet{EtaCar:Pittard02}). The total
kinetic energy in stellar winds is of the order of a few
$10^{37}\UNITS{erg\,s^{-1}}$ for the LBV and the OB star together.

When stellar winds of such stars collide, they form a stellar wind
shock, where particles can be accelerated to non-thermal energies
\citep[e.g.][]{CWB:Eichler,CWB:Reimer06}. There is strong evidence for
the existence of non-thermal particles in \etacar\ based on X-ray
measurements performed with the instruments aboard the
\textit{INTEGRAL} \citep{EtaCar:Integral,EtaCar:Leyder10} and
\emph{Suzaku} satellites \citep{EtaCar:Suzaku09}. In the high-energy
(HE; $100\UNITS{MeV}\leq E\leq 100\UNITS{GeV}$) domain, the
\textit{AGILE} \citep{EtaCar:Agile} and \textit{Fermi}-LAT
\citep{FERMI:BSL,EtaCar:Fermi10,FERMI:1yr,Fermi:2yr} collaborations
have reported on the detection of a source coincident with \etacar\
(henceforth \TwoFGL). Recently \citet{EtaCar:Farnier11} confirmed with
the \textit{Fermi}-LAT data the position of the HE \g-ray source and
extracted an energy spectrum which features a low and a high-energy
component. The high-energy component extends up to $\sim100$\,GeV,
close to the energy threshold of the \hess\ telescope array. The
\textit{AGILE} collaboration reported on a two-day \g-ray flare from
the direction of \etacar\ which occurred in October 2008. Although
this increased \g-ray flux could not be confirmed by
\citet{EtaCar:Farnier11}, \citet{EtaCar:Walter11} found that the
high-energy component flux shows a drop in the yearly light
curve. Both these findings point to a possible origin of the HE \g-ray
emission in the colliding wind region of \etacar.

TeV~J2032+4130 \citep{HEGRA:Cygnus}, HESS~J1023--575 \citep{HESS:Wd2}
and the extended VHE \g-ray emission seen from the vicinity of
Westerlund~1 \citep{HESS:Wd1_11} seem to indicate that VHE \g-ray
emission can be linked to massive stars in our Galaxy and motivates an
investigation of \etacar\ and the Carina region as a whole as
potential VHE \g-ray emitters. A further motivation comes from the
detection of \g-ray emission from binary star systems such as LS\,5039
\citep{HESS:LS5039}, PSR\,B1259--63 \citep{HESS:1259}, LS\,I\,+61\,303
\citep{MAGIC:LSI} and the probable TeV binary HESS\,J0632+057
\citep{HESS:Monoceros,0632:Bongiorno11}. Note that, unlike \etacar\,
these objects have a compact object (a neutron star or black hole) as
stellar companion. Furthermore, the recent detection of HE \g-ray
emission up to $100\UNITS{GeV}$ from the direction of \etacar\ might
hint at particle acceleration up to the VHE \g-ray regime in which
\hess\ is operating.

\section{\hess\ observations}\label{sec:HESS_OBS}
\subsection{\hess\ Experiment}

\hess\ is an array of four VHE \g-ray imaging atmospheric Cherenkov
telescopes (IACTs) located in the Khomas Highland of Namibia. Each of
these telescopes is equipped with a tessellated spherical mirror of
107\,m$^2$ area and a camera comprising 960 photomultiplier tubes,
covering a large Field-of-View (FoV) of 5$^{\circ}$ diameter. The
system works in a coincidence mode, requiring at least two of the four
telescopes to detect the same extended air shower. This stereoscopic
approach results in an angular resolution of $\sim6'$ per event, a
good energy resolution (15\% on average) and an efficient rejection of
the hadronic background \citep[selection cuts retain less than 0.01\%
of the CRs;][]{Benbow:Gamma04}. \hess\ has a point-source sensitivity
of $\sim2\times10^{-13}\UNITS{ph\,cm^{-2}\,s^{-1}}$ within 25 hours of
observations \citep{HESS:Crab}. This flux level corresponds to a 1\%
integral flux of the Crab Nebula for energies $E\,>\,$0.2\,TeV, and
detection threshold of 5$\sigma$ \citep{LiMa}. The more advanced data
analysis method that is used in this work is discussed later, and
achieves a significantly better point-source sensitivity \citep{TMVA}.

\subsection{Data Set}
Observations of the (Sagittarius-) Carina arm tangent have been
carried out as part of the \hess\ Galactic plane survey
\citep{HESS:GPS06,HESS:DarkSources}. Additionally, observations
pointing in the direction of \etacar\ have been performed in the
so-called \emph{wobble-mode}, where the telescopes were alternately
pointed offset in RA and Dec from \etacar\ \citep{HESS:Crab}. The
Carina region and its surroundings were observed with the \hess\ array
for a total of 62.4\,hours between 2004 and 2010. After standard data
quality selection, where data taken under unstable weather conditions
or with malfunctioning hardware have been excluded, the total exposure
time after dead time correction of 3 to 4 telescope data is 33.1 hours
\citep{HESS:Crab}. Due to \etacar's very southern position on the sky,
observations have been carried out at moderate zenith angles of
36\degree\ to 54\degree, with a mean value of 39\degree. The average
pointing offset from \etacar\ was 0.8\degree.

\subsection{Data Analysis}
The available data have been analysed with the \hess\ Standard
Analysis for shower reconstruction \citep{HESS:Crab} and the
Hillas-based Boosted Decision Trees (BDT) method for an efficient
suppression of the hadronic background component \footnote{The
  \emph{HESS Analysis Package (HAP)} version 11-02-pl07 has been used
  to analyse the data set presented in this work.}. This machine
learning algorithm returns a continuous variable (called $\zeta$) that
was used to select \g-ray-like events. Compared to the \hess\ Standard
Analysis, a cut on this parameter results in an improvement in terms
of sensitivity of $\sim20$\% for spectral and morphological analysis
For the generation of sky images, the spectral analysis and the
production of light curves, the \emph{\zetastd-cuts} with a 60
photoelectron (p.e.) cut on the image intensity has been applied
\citep[see][]{TMVA}. The usage of this set of cuts leads to an energy
threshold of 470\,GeV for these observations. The 68\% containment
radius of the \hess\ point spread function (PSF) for the analysis
presented here is $6.7'$.

In order to search for a \g-ray signal from \etacar\ and the Carina
Nebula, two different background estimation techniques have been
employed, i.e. the \emph{ring} background and the \emph{reflected}
background model \citep{HESS:Background}. The former has been applied
to produce two-dimensional sky images, whereas the latter method has
been used to derive spectral information and light
curves. Table~\ref{tab:eta_datasets} summarises the properties of the
different data sets used in this work and the orbital phases of
\etacar\ which are covered by \hess\ observations. Note that
throughout the paper the orbital phase is defined as phase angle with
reference zero-time MJD 52822.492 corresponding to the periastron
passage, and a period of 2022.7 days \citep{EtaCar:Damineli08}.

\begin{table*}
  \centering
  \caption{Properties of the data sets used to calculate the flux upper limits 
    and the light curve shown in Fig.~\ref{fig:EtaCar_stdzeta_ULs} 
    and \ref{fig:EtaCar_stdzeta_MWL_LC}, respectively. The time range
    of the \hess\ observations, the total live time corresponding to the individual data sets
    along with the covered orbital phase are summarised.}
  \begin{tabular}{@{}cccccc}
    \hline
    Data Set & date & MJD & Phase & live time (hrs) \\ \hline
    1 & 24.03.04 & 53088 & 0.12 & 1.4 \\
    2 & 11.02.05 & 53412 & 0.29 & 0.9 \\
    3 & 22.05.06 - 24.05.06 & 53877 - 53879 & 0.52 & 5.0 \\
    4 & 01.02.09 - 14.04.09 & 54863 - 54935 & 1.01 - 1.04 & 1.8 \\
    5 & 15.01.10 - 22.03.10 & 55211 - 55277 & 1.18 - 1.21 & 18.4 \\
    6 & 06.12.10 - 18.12.10 & 55536 - 55548 & 1.34 - 1.35 & 5.6 \\ \hline
    all & 24.03.04 - 18.12.10 & 53088 - 55548 & 0.12 - 1.35 & 33.1 \\ \hline
  \end{tabular}
  \label{tab:eta_datasets}
\end{table*}

Observations have been carried out over a time span of six years,
during which the reflectivity of the \hess\ mirrors varied and the
gains of the photomultipliers (PMTs) changed. The energy scale of the
instrument is calibrated by looking at the response to single muons
\citep{HESS:Crab}.

Two different circular regions have been selected a priori and have
been searched for a signal in the \hess\ data. Both of them are shown
in Fig.~\ref{fig:EtaCar_stdzeta_sign} and are centred on the \etacar\
position at RA \HMS{10}{45}{03.6} and Dec \DMS{-59}{41}{04.3}
(J2000). Given the size of the \etacar\ system of \order($1'$), any
VHE \g-ray signal would appear point-like to \hess\ (\emph{Region 1},
$0.112^\circ$ radius). The Carina Nebula, on the other hand, is a
large and complex reflection nebula which shows extended emission seen
in mid-infrared, optical and X-ray wavelengths on scales of
$\sim1^\circ\times2.5^\circ$. The second circular region (\emph{Region
  2}, $0.4^\circ$ radius) has a physical scale of 16\,pc at 2.3\,kpc
distance and has been chosen such that the bulk of the diffuse X-ray
emission \citep{Car:Townsley11} and potential particle acceleration
sites such as the massive young stellar clusters Tr\,14, and Tr\,16
are encompassed. \emph{Region 2} also encloses most of the H$\alpha$
\citep{Car:Muller98} and $8\,\mu$m emission which traces gaseous and
dusty material.

All results presented in the following have been successfully checked
for consistency with an analysis chain that is based on a different
shower reconstruction method and \g-ray selection criteria
\citep{Model++}, and on a different calibration. During data taking,
increased and variable single-telescope rates and, after quality
selection, an increased but stable system trigger rate have been
observed. This can be ascribed to the very high night-sky-background
(NSB) level caused by the strong UV emission from the Carina
Nebula. This NSB level is higher than in any other \hess\ FoV from
which results have been reported so far. Systematic tests have been
performed and show that predominantly events which result in shower
images with intensities below 60\,p.e. are affected. However, the high
NSB level does not affect the results presented here, since only
events with image sizes greater than 60\,p.e. are used. Moreover, the
main analysis and the cross-check analysis -- which models the NSB for
shower reconstruction \citep{Model++} -- give consistent results.

\subsection{VHE \g-ray results}

\begin{figure}
  \centering
  \resizebox{\hsize}{!}{\includegraphics[]{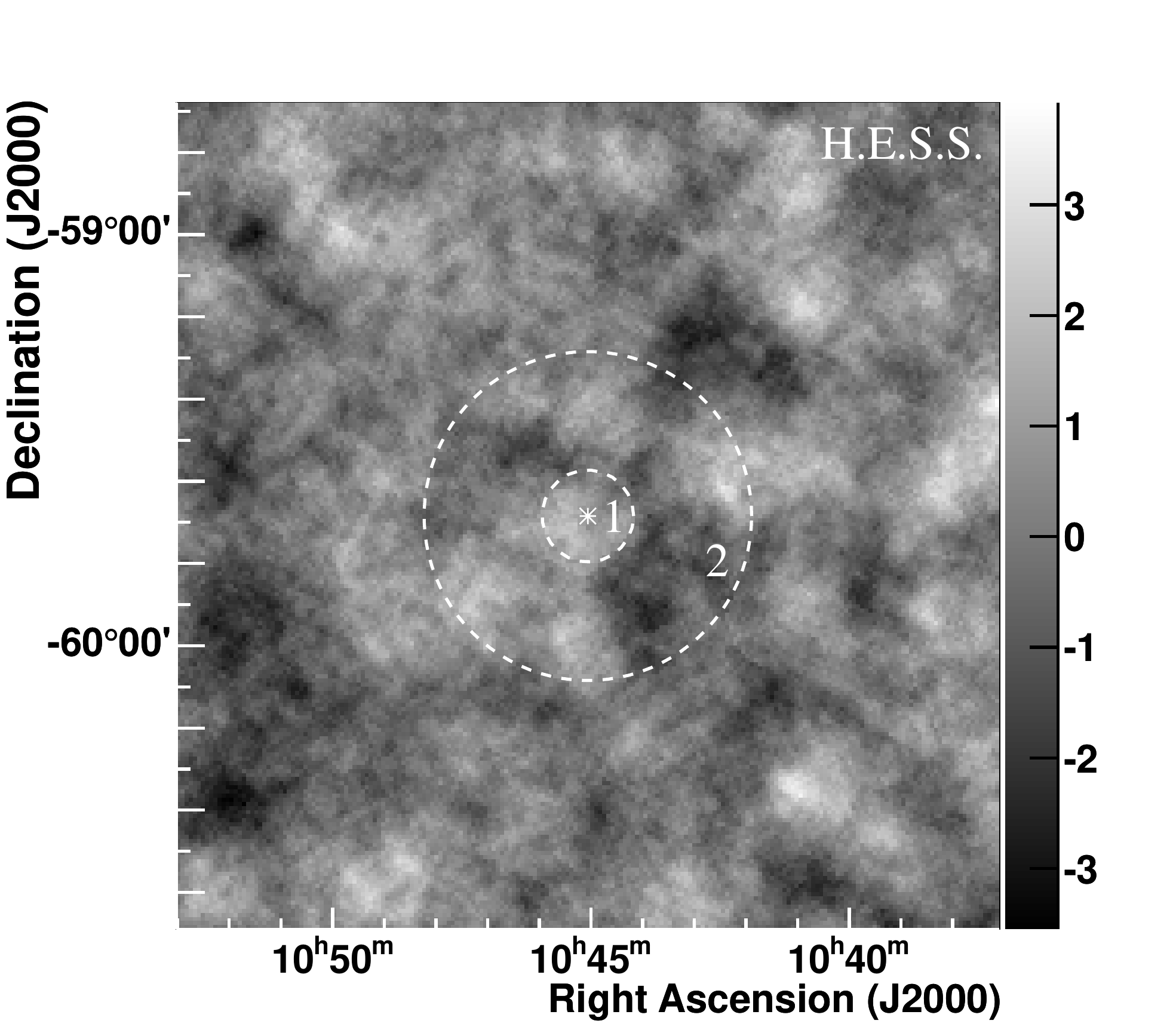}}
  \caption{\hess\ significance map of the $2^\circ\times 2^\circ$
    region around \etacar, generated with an oversampling radius of
    $6.7'$, corresponding to the PSF of this analysis, and calculated
    according to \citet{LiMa}. Circles denote the integration radii
    (\emph{Region 1}\&\emph{2}) which were used to extract the
    statistics as given in the text and the upper limits depicted in
    Fig.~\ref{fig:EtaCar_stdzeta_ULs}.}
\label{fig:EtaCar_stdzeta_sign}
\end{figure}

Fig.~\ref{fig:EtaCar_stdzeta_sign} shows the VHE \g-ray significance
map of the $2^\circ\times2^\circ$ region centred on the optical
position of \etacar, and calculated according to \citet{LiMa}. The map
has been obtained with the \emph{ring} background method and for an
integration angle of $6.7'$. No evidence for significant VHE \g-ray
emission is found from \emph{Region 1} or from \emph{Region
  2}. Assuming a point-like source at the position of \etacar\ ({\em
  Region 1}) a total of $40\pm 26$ excess events with a significance
of $1.6\sigma$ are found. Within \emph{Region 2}, $197\pm 101$ excess
events with a significance of $2.0\sigma$ are detected.

\begin{figure}
  \centering 
  \resizebox{\hsize}{!}{\includegraphics[]{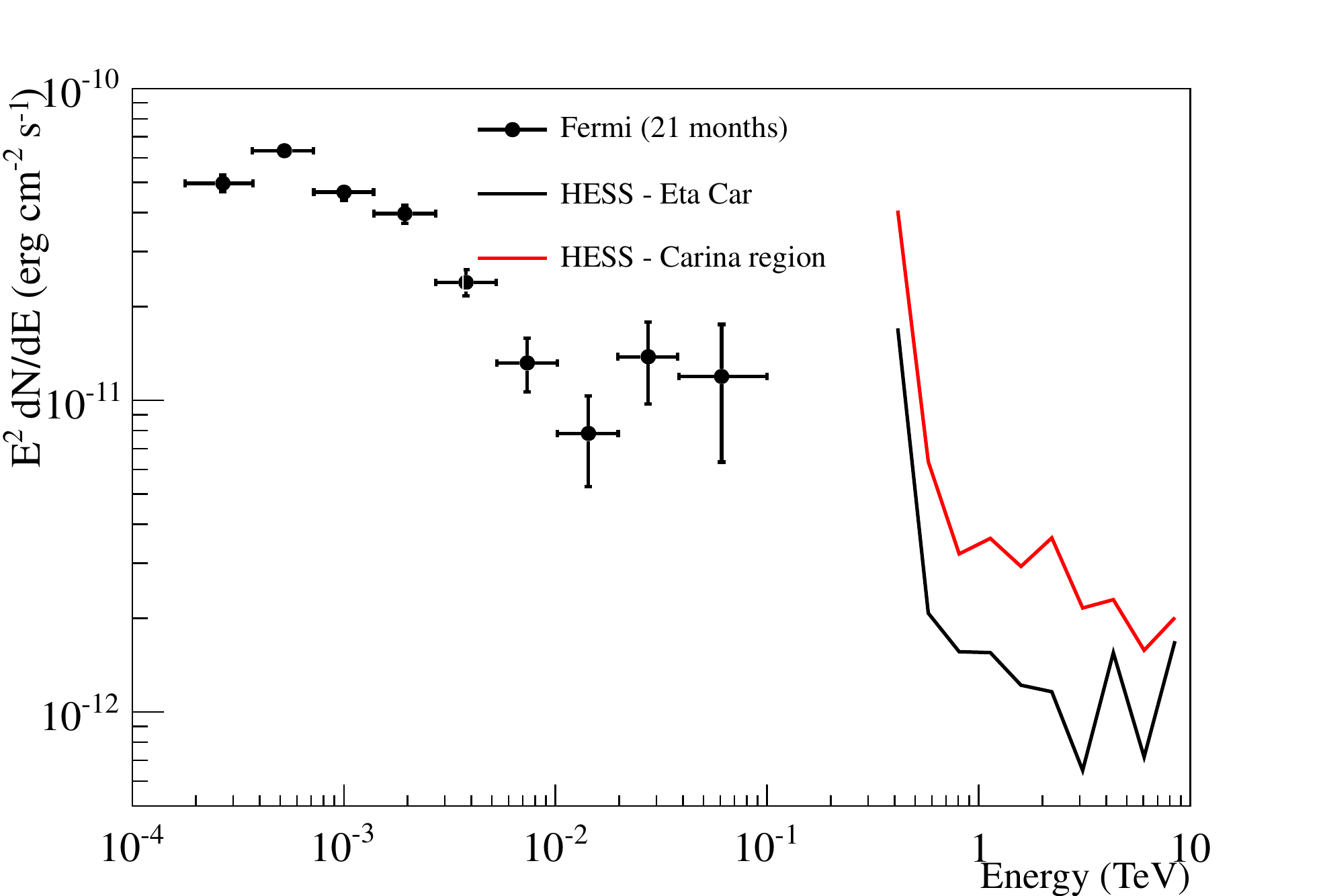}}
  \caption{\hess\ upper limits (99\% confidence level) on the VHE
    \g-ray flux from \etacar\ and the Carina Nebula. Also shown are
    the spectral points for \TwoFGL\ as derived by
    \citet{EtaCar:Farnier11}. Note that the \emph{Fermi}-LAT spectrum
    and the \hess\ ULs have been obtained from data which were not
    taken contemporarily. Due to the lack of statistics, ULs at higher
    energies could not be produced.}
\label{fig:EtaCar_stdzeta_ULs}
\end{figure}

Upper limits (ULs) for the VHE \g-ray emission from \etacar\ and the
extended region of 0.4\degree\ radius which covers the inner parts of
the Carina Nebula have been
produced. Fig.~\ref{fig:EtaCar_stdzeta_ULs} shows the 99\% ULs
\citep[following][]{FeldmanCousins98} on the VHE \g-ray flux from
\etacar\ and the Carina region, assuming an underlying power law
distribution $dN/dE=\Phi_0\cdot(E/1\UNITS{TeV})^{-\Gamma}$ with photon
index $\Gamma=2.0$. Adjusting the assumed spectral index to
$\Gamma=2.5$ changes the presented upper limits by less than 2\%. Also
shown is the HE \g-ray flux from the point-like source \TwoFGL,
coincident with \etacar, as detected by the LAT instrument onboard the
\textit{Fermi} satellite
\citep{FERMI:BSL,FERMI:1yr,EtaCar:Farnier11}. Above the energy
threshold of 470\,GeV, the derived 99\% integral flux ULs are
$7.7\times10^{-13}\UNITS{ph\,cm^{-2}\,s^{-1}}$ for a point-like source
at the position of \etacar\ and
$4.2\times10^{-12}\UNITS{ph\,cm^{-2}\,s^{-1}}$ for the extended
\emph{Region 2}.

\begin{figure*}
  \centering
  \resizebox{\hsize}{!}{\includegraphics[]{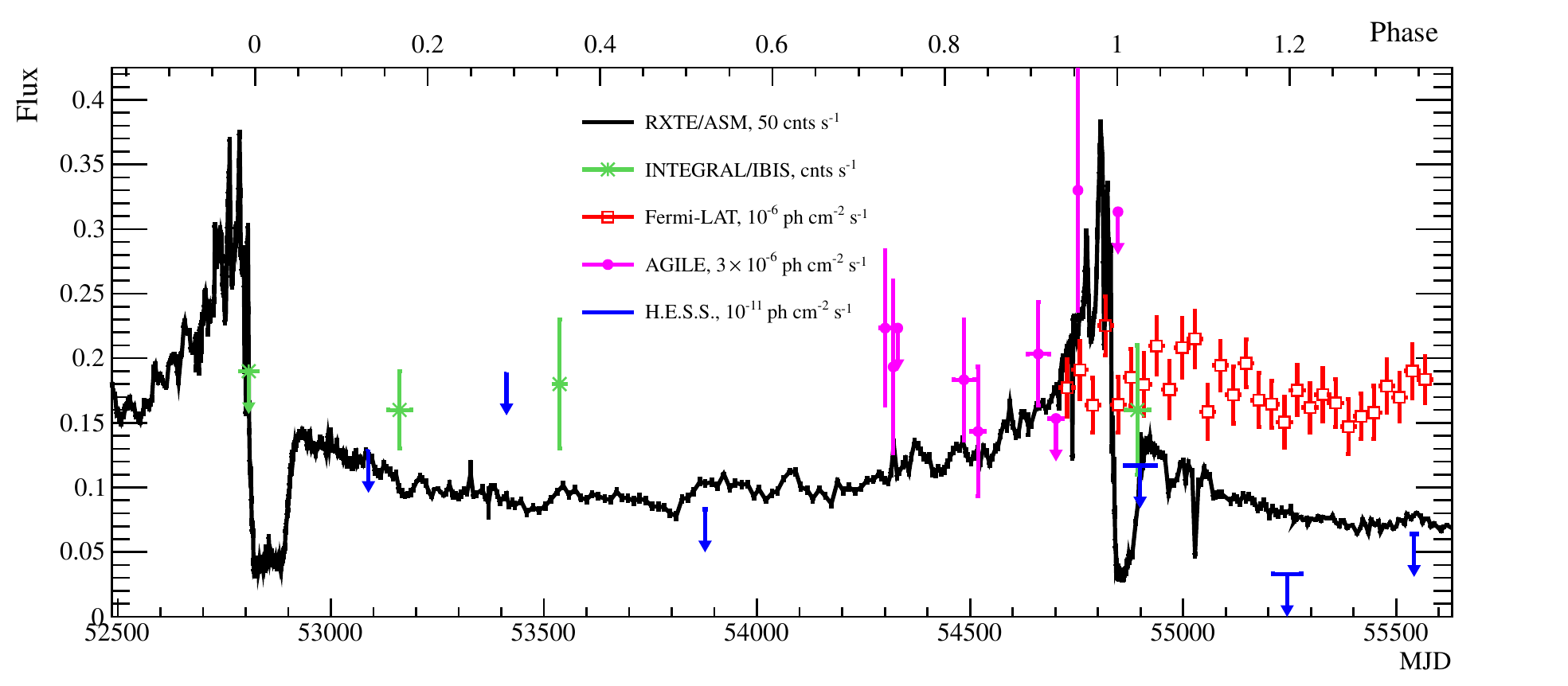}}
  \caption{\hess\ flux upper limits (99\% confidence level) for
    \etacar\ and the six data sets described in
    Table~\ref{tab:eta_datasets} (blue). Also shown are the
    \textit{RXTE/ASM} light curve \citep[black, 50 cnts/s correspond
    to $\sim3.75\times
    10^{-10}$\,erg\,cm$^{-2}$\,s$^{-1}$,][]{EtaCar:Corcoran10},
    \textit{INTEGRAL/IBIS} data points \citep[green, 0.15 cnts/s
    correspond to $\sim1.11\times
    10^{-11}$\,erg\,cm$^{-2}$\,s$^{-1}$,][]{EtaCar:Integral},
    \emph{AGILE} measurements \citep[purple,][]{EtaCar:Agile} and the
    monthly \textit{Fermi}-LAT light curve (red). The \emph{AGILE}
    flare is not shown, but it falls around MJD 54753 with a flux of
    $0.90\pm0.22$ in this representation. Note that the x-errors on
    flux points and ULs indicate the different time periods covered by
    observations of the different instruments
    (e.g. Table~\ref{tab:eta_datasets}).}
\label{fig:EtaCar_stdzeta_MWL_LC}
\end{figure*}

The light curve of the binary system \etacar\ shows variability in the
optical \citep[e.g.][]{EtaCar:Damineli00}, IR
\citep[e.g.][]{EtaCar:Whitelock04} X-ray \citep{EtaCar:Corcoran10} and
HE \g-ray band \citep{EtaCar:Walter11} on timescales of months to
years. In order to search for a possible variability in VHE \g\ rays
on similar timescales, the data collected during the \hess\
observations between 2004 and 2010 have been split into six different
data sets accordingly (see Table~\ref{tab:eta_datasets}). Since no VHE
\g-ray signal could be found in any of these data sets, flux ULs have
been derived for the covered time periods using the same assumptions
as before. The statistics, energy thresholds and ULs are summarised in
Table~\ref{tab:eta_stats}. Fig.~\ref{fig:EtaCar_stdzeta_MWL_LC} shows
the \hess\ flux ULs (99\% confidence level) above $1\UNITS{TeV}$ at
the different orbital phases of \etacar. Also shown are the
\textit{RXTE/ASM} light curve and the \textit{INTEGRAL/IBIS} data
points in the X-ray domain as well as the \emph{AGILE} and monthly
\textit{Fermi}-LAT light curve in HE \g\ rays \footnote{The light
  curve has been obtained following the procedure described in
  \citet{EtaCar:Farnier11}, but for an extended data set of 30 months
  (MJD 54682 to MJD 55595)}.
\begin{table*}
  \caption{Statistics and flux upper limits for the \hess\ \etacar\ data sets.}
  \begin{tabular}{@{}cccccccccc}
    \hline
    Data Set & \emph{On} & \emph{Off} & $\alpha$ & Excess & Significance & $E_{\mathrm{th}}$ & $F_{99}(>E_{\mathrm{th}})$ & 
    $F_{99}(>$1\,TeV) & Phase\\
    & & & & & $\sigma$ & TeV & $10^{-12}$\,ph\,cm$^{-2}$\,s$^{-1}$ &
    $10^{-12}$\,ph\,cm$^{-2}$\,s$^{-1}$ & \\\hline
    1 & 70 & 414 & 0.1633 & 2.4 & 0.3 & 0.43 & 2.99 & 1.29 & 0.12\\
    2 & 14 & 218 & 0.0543 & 2.2 & 0.6 & 0.43 & 4.46 & 1.89 & 0.29\\
    3 & 85 & 2300& 0.0383 & -3.2& -0.3& 0.47 & 1.78 & 0.83 & 0.52\\
    4 & 29 & 236 & 0.0875 & 8.3 & 1.6 & 0.52 & 2.26 & 1.17 & 1.01-1.04\\
    5 & 350 & 3744 & 0.0852 & 31.1& 1.6 & 0.52 & 0.64 & 0.33 & 1.18-1.21\\
    6 & 100 & 2364 & 0.0426 & -0.6& -0.1& 0.52 & 1.23 & 0.64 & 1.34-1.35\\ \hline
    all & 648 & 11248 & 0.0540 & 40.2 & 1.6 & 0.47 & 0.77 & 0.36 & 0.12 - 1.35 \\\hline
  \end{tabular}

  \emph{On} denotes the number of \g-ray-like events from \emph{Region 1}, 
  \emph{Off} the number of \g-ray-like events from the background control regions, 
  $\alpha$ is the normalisation factor between the \emph{On} and \emph{Off} exposures, 
  $E_{\mathrm{th}}$ is the energy threshold in TeV, and $F_{99}(>E_{\mathrm{th}})$ and 
  $F_{99}(>$1\,TeV) are the 99\% flux ULs above $E_{\mathrm{th}}$ and 1~TeV, 
  respectively, following \citet{FeldmanCousins98}.
  \label{tab:eta_stats}
\end{table*}

\section{Discussion}
\label{sec:discussion}

\subsection{\etacar}

The detection of point-like HE \g-ray emission from \TwoFGL\ was
originally reported in the 3-months bright source list
\citep{FERMI:BSL} and was confirmed by \citet{EtaCar:Farnier11} based
on 21 months of data. The spectrum presented by
\citet{EtaCar:Farnier11} shows two distinct features: a low-energy
component which is best fitted by a power law with index
$\Gamma=1.69\pm 0.12$ and exponential cut-off at $1.8\pm0.5$\,GeV and
a high-energy component which extends to $\sim100$\,GeV and is well
described by a simple power law with index $1.85\pm0.25$. If the HE
\g-ray flux shown in Fig.~\ref{fig:EtaCar_stdzeta_ULs} extended to the
TeV regime, it would have been detectable in the \hess\ data presented
in this work. The non-detection of a significant VHE \g-ray signal
from \etacar\ at any orbital phase and in the complete \hess\ data set
has some interesting implications for the origin of the HE \g-ray
emission which are discussed below.

\citet{EtaCar:Walter11} showed that the flux of the high-energy
component (E~$>$~10~GeV) decreases by a factor of 2--3 in the yearly
light curve, which could point to a scenario in which the parent
particle population is accelerated in the colliding wind region of the
binary system
\citep{EtaCar:Agile,EtaCar:Farnier11,EtaCar:Bednarek11}. However, the
low-energy component does not seem to vary on yearly or monthly
timescales. For the colliding wind model, the lower energy component
($0.2\,\mathrm{GeV}\le E \le 10$\,GeV) detected by the LAT is
interpreted as Inverse Compton (IC) \g-ray emission produced in
interactions of the accelerated electrons with the dense stellar
radiation fields of the binary stars. The hard HE \g-ray component can
be interpreted in the colliding wind region model as either
$\pi^0$-decay \g\ rays, which are produced in proton-proton
interactions in the dense stellar wind material \citep[][their Model
B]{EtaCar:Farnier11,EtaCar:Bednarek11} or as a second leptonic IC
contribution \citep[Model A in][]{EtaCar:Bednarek11}. Interestingly,
the \hess\ flux ULs for the individual subsets above the threshold
energies of $\sim0.5$\,TeV are all well below the extrapolated hard HE
\g-ray component measured by \textit{Fermi}-LAT (which is at a level
of $\sim1\times 10^{-11}\UNITS{erg\,cm^{-2}\,s^{-1}}$) \footnote{Note
  that for a steeper spectral index of the high-energy LAT component,
  i.e. $\Gamma \gtrsim 2.5$, the H.E.S.S. ULs of the individual data
  sets are compatible with the \emph{Fermi}-LAT spectrum.}. This
implies that the \g\ radiation spectrum has a cut-off below
$\sim1\UNITS{TeV}$, caused either by a cut-off in the accelerated
particle spectrum or resulting from significant \g-\g\ absorption in
the radiation field close to the two stars in the colliding wind
region model. \citet{EtaCar:Bednarek11} concluded that in the case of
accelerated protons, the resulting $\pi^0$-decay \g-ray emission
should extend to TeV energies at phases far from periastron. The
\hess\ data do not show \g-ray emission in the multi-TeV range at any
orbital phase. Note, however, that the maximum detectable photon
energy critically depends on the \g-\g\ absorption at the location
where the photon is emitted, and on the alignment between the \g-ray
production region, the star and the observer. For \etacar, the optical
depth for TeV particles becomes smaller than unity only at phases far
from periastron, where the radiation field densities of both
stars are low enough to allow \g\ rays to escape the system \citep[see
e.g. Fig.~3 in][]{EtaCar:Bednarek11}.

In an alternative scenario, particles are assumed to be accelerated in
the outer blast wave which originates in the Great Eruption
\citep{EtaCar:Ohm10}. For a potential non-variable hadronic
high-energy \g-ray component, as discussed in
\citet{EtaCar:Skilton12}, the maximum particle energy of the parent
proton population is limited by three different parameters: the time
since the giant outburst, i.e. $167\UNITS{yrs}$, the blast wave speed,
which is measured as $3500-6000$\,km\,s$^{-1}$ \citep{EtaCar:Smith08}
and the magnetic field, which is only poorly constrained. Contrary to
the effects close to the wind-wind collision region, \g-\g\ absorption
at the location of the blast wave has no significant effect on the
\g-ray spectrum, given that the optical depth $\tau_{\gamma\gamma}$ at
this location is orders of magnitude smaller than in the colliding
wind region. For the parameters used in \citet{EtaCar:Ohm10}
\footnote{Note that \citet{EtaCar:Ohm10} work in the limit of Bohm
  diffusion which might overestimate the particle acceleration
  efficiency.}, the maximum energy of protons producing \g\ rays of
$0.5-1.0$\,TeV energy would be of \order$(5-10)\UNITS{TeV}$ for
magnetic field strengths in the blast wave of $(3-10)\UNITS{\mu G}$
and a blast wave speed of $3500\UNITS{km\,s^{-1}}$. For this set of
parameters, the \hess\ measurement excludes larger magnetic fields
and/or higher blast wave speeds for this model.

\subsection{Carina Nebula}

The Carina Nebula harbours many potential particle acceleration sites
such as massive binary systems \citep[e.g. WR~25, \etacar, or the
recently discovered HD~93250,][]{Sana11}, young massive stellar
clusters (e.g. Tr 14, Tr 16) and possibly one or more SNR
shells. Electrons and hadrons accelerated at these places would
diffuse out of the acceleration region and interact with interstellar
radiation fields and/or gaseous material, producing \g-ray emission
via $\pi^0$-decay or IC processes. Potential HE or VHE \g-ray emission
could therefore trace the regions where a SNR shell interacts with
high-density gas in molecular clouds (MCs) \citep[as observed e.g. for
W~28][]{HESS:W28,Fermi:W28}. Additionally, low-energy CRs could be
traced by ionisation of MCs \citep[see e.g.][]{Ceccarelli2011}.

\citet{Car:Townsley11} investigated the complex structure and
composition of the diffuse X-ray emission in the Carina Nebula with
the \emph{Chandra} satellite. The spectrum of this emission is
phenomenologically best described by a multi-component model of
different thermal plasmas in collisional ionisation equilibrium and in
a non-equilibrium ionisation state. The X-ray emission does not seem
to show any hint of a non-thermal component which would be indicative
of particle acceleration in this region. Possible explanations for
these observations are: e.g. that currently no particle acceleration
is taking place and hence non-thermal emission is not expected, or
that the potential synchrotron emission has a much lower flux level
than the efficient plasma emission, or the SNR shock has been diluted
in the ambient plasma. If, however, particle acceleration occurred in
the past at e.g. the shocks from one or more potential SNR shells,
electrons might have cooled via synchrotron or IC radiation to a level
not detectable by \emph{Chandra} or below the \hess\ UL, respectively.
Note that for a far-infrared luminosity of $L_{\mathrm{Car}}
\sim7\times10^{6}L_\odot$ \citep{Car:Salatino12} and a circular region
of 16\,pc radius, the IC cooling time for 1\,TeV electrons would be
$\tau_{\mathrm{IC}} \sim6 \times 10^3 \UNITS{yrs}$. 

CR hadrons on the other hand diffuse out of the acceleration region
and interact with the gaseous or dusty material, producing
$\pi^0$-decay \g\ rays.  The \hess\ ULs can be used to constrain the
CR density enhancement factor $\kappa_{\mathrm{CR}}$ in units of the
local CR density using Equation (10) from \citet{Aharonian91},
assuming that all the gas located in \emph{Region 2} is irradiated by
CRs at the same time. Following \citet{Car:Preibisch2011b} and
\citet{Car:Yonekura2005}, the total gas and dust mass in \emph{Region
  2} can be estimated to $\sim1.5 \times 10^5 \,M_\odot$. At a
distance of 2.3\,kpc this gives $\kappa_{\mathrm CR} = 23/f$, where
$f$ is the fraction of the molecular cloud mass effectively irradiated
by high-energy CRs. Assuming $f = 1$, this value can be compared to
the CR enhancement factors obtained from the \hess\ detection of VHE
$\gamma$-ray emission from W~28 \citep{HESS:W28}. W~28 is an old
$((3.5-15) \times 10^4$ yrs, Kaspi et al. 1993), mixed-morphology SNR,
which is seen to interact with molecular clouds belonging to the same
massive star forming region \citep[e.g.][]{Brogan06}. \citet{HESS:W28}
derive $\kappa_{\mathrm CR}(\mathrm{W~28}) = 13-32$ for clouds with
masses $(0.2-1.5) \times 10^5\,M_\odot$ and distances between 2\,kpc
and 4\,kpc. However, there is at present no evidence for a SNR in the
Carina Nebula, although SN explosions must have already occurred in
the past (say $\sim10^6$\,yrs ago), in view of the presence of a
neutron star. In that case, the lack of GeV-TeV emission from the
nebula may have two explanations, separate or combined: (i) the
factor $f$ being $\ll1$ due to diffusive or advective transport of CRs
in the region (too slow to fill the region or so fast that they
escape), in which case the upper limit to $k_{\mathrm{CR}} \gg
k_{\mathrm{CR}} (\mathrm{W~28})$, and/or (ii) the $p-p$ collision
timescale (for an average gas density of $100-400$\,cm$^{-3}$ in the
50\,pc region) is about 10 times less than the age of putative SNRs.

\section{Summary}

The search for VHE \g-ray emission from the colliding wind binary
\etacar\ and the most active HII region in the Galaxy, the Carina
Nebula, has been presented. No sign of VHE \g-ray emission could be
detected by \hess\ for \etacar\ and a 99\% UL on the integral \g-ray
flux of $7.7\times10^{-13}\UNITS{ph\,cm^{-2}\,s^{-1}}$ above 470\,GeV
has been derived using a 33-hour data set collected over 6 years and
covering the full phase range of the binary. Given the detection of a
HE \g-ray component by \textit{Fermi}-LAT, that extents up to
$\sim100$\,GeV, and assuming a spectral index of the high-energy
\emph{Fermi}-LAT component as found for the average spectrum by
\citet{EtaCar:Farnier11}, the derived \hess\ ULs imply a cut-off in
the \g-ray spectrum below a few hundred GeV. \hess\ observations did
not reveal significant VHE \g-ray emission from the Carina Nebula
either. The derived ULs allow us to estimate the CR enhancement factor
in this region ($<$ 23) which is at a comparable level to the values
obtained for the W~28 complex, assuming that CRs illuminate the whole
cloud complex. H.E.S.S.~II, which adds a 600\,m$^2$ telescope to the
existing system, will be operational during the next periastron
passage in mid 2014 and will be sensitive to lower energies. Together
with the future Cherenkov Telescope Array \citep[CTA,][]{CTA:Actis11},
with its greatly improved sensitivity and broader energy coverage,
both instruments will close the gap between the HE and VHE \g-ray
range and will allow to probe the cut-off region in the \g-ray
spectrum of \etacar\ and to search for any variability in this system
at very high energies.

\section*{Acknowledgements} 
We thank the referee R. Walter for his helpful comments and
suggestions. The support of the Namibian authorities and of the
University of Namibia in facilitating the construction and operation
of H.E.S.S. is gratefully acknowledged, as is the support by the
German Ministry for Education and Research (BMBF), the Max Planck
Society, the German Research Foundation (DFG), the French Ministry for
Research, the CNRS-IN2P3 and the Astroparticle Interdisciplinary
Programme of the CNRS, the U.K. Science and Technology Facilities
Council (STFC), the IPNP of the Charles University, the Czech Science
Foundation, the Polish Ministry of Science and Higher Education, the
South African Department of Science and Technology and National
Research Foundation, and by the University of Namibia. We appreciate
the excellent work of the technical support staff in Berlin, Durham,
Hamburg, Heidelberg, Palaiseau, Paris, Saclay, and in Namibia in the
construction and operation of the equipment. SO acknowledges the
support of the Humboldt foundation by a Feodor-Lynen research
fellowship.

\bibliographystyle{mn2e_williams}
\bibliography{HESS_EtaCar_v2}
\label{lastpage}

\end{document}